\documentclass[a4paper,11pt]{article}
\usepackage{amsmath,amssymb}
\def\be{\begin{equation}}
\def\ee{\end{equation}}

\def\f{\frac}

\def\bea{\begin{eqnarray}}

\def\eea{\end{eqnarray}}

\def\ll{\langle\langle}
\def\rr{\rangle\rangle}
\def\rot{\mathop{\mathrm{rot}}}
\def\grad{\mathop{\mathrm{grad}}}
\def\div{\mathop{\mathrm{div}}}

\def\ben{\begin{displaymath}}
\def\een{\end{displaymath}}
\def\ba{\begin{array}{c}}
\def\bal{\begin{array}{l}}
\def\ea{\end{array}}
\def\p{\partial}

\begin{document}


\vspace{1.5cm}

 \begin{center}{\Large \bf
 A Positive-Definite Scalar Product for Free Proca Particle
  }\end{center}

\vspace{10mm}

 \begin{center}

 {\bf V\'{\i}t Jakubsk\'{y}}

 \vspace{3mm}

 \'{U}stav jadern\'e fyziky AV \v{C}R

250 68 \v{R}e\v{z}

Czech Republic

{e-mail: jakub@ujf.cas.cz}

\vspace{5mm}

{\bf Jaroslav Smejkal}

\vspace{3mm}

\' Ustav technick\'e a experiment\'aln\'i fyziky \v CVUT

   Horsk\'a 3a/22\\                                         
     128 00 Praha 2 - Nov\'e M\v esto. 
Czech Republic

{e-mail: Jaroslav.Smejkal@utef.cvut.cz}

\vspace{3mm}
%


\vspace{5mm}

\end{center}

\vspace{5mm}

\section*{Abstract}
We implement recent results of pseudo-Hermitian quantum mechanics to description of relativistic massive particle with spin-one. We derive a one-parameter family of Lorentz invariant positive-definite scalar products on the space of solutions of Proca equation.


\section{Introduction}

As the non-relativistic quantum theory established to be a consistent physical theory, a natural task of intertwining the quantum mechanics with relativistic theory raised. 
The first attempts to match these two theories has led to a relativistic quantum kinematics governed by Klein-Gordon, Dirac and Proca equations \cite{KGDirac}\footnote{Hierarchy of covariant equations of higher spin has been found, see \cite{landau}. For our present purposes, we will consider system with spin $s\leq 1$.}. A manifestly covariant equation describing a free massive vector boson 
 
\be
  \partial_{\mu}F^{\mu\nu}+m^2A^{\nu}=0,\ \ \ \ \
 F^{\mu\nu}=\partial^{\mu}A^{\nu}-\partial^{\nu}A^{\mu}.
 \label{1}
 \ee
has been derived by Proca \cite{KGDirac}. Similarly to its spin-zero counterpart, Klein-Gordon equation, it contains a second time derivative. This has a deep consequences resulting in the well-known puzzle of probabilistic interpretation of the associated bosonic theories.

In the forties of the previous century, the formalism of the relativistic equations has been unified formally (let us mention works of Sakata and Taketani \cite{ST}, Foldy \cite{foldy} and Feschbach and Villars \cite{FV} in this context). Similarly to Dirac equation, evolution equation (\ref{1}) was rewritten as $i\partial_t\Psi=H_{ST}\Psi$. The sequential lost of its manifest covariance was balanced by improved description of spin-one particles and anti-particles as the wave function consists of six components. On the contrary to Dirac equation, Hamiltonian represented by $6\times 6$ matrix 

 \be H_{ST}=(\tau_3+i\tau_2)\frac{{\bf P}^2}{2m}+\tau_3 m-i\tau_2\frac{({\bf SP})^2}{m}, \label{ST}\ee

\noindent
ceases to be Hermitian. Indeed, it rather fulfills the relation
\footnote{The matrices $S_j$ constitute three dimensional representation of rotational group, see (\ref{reprrot}), $\tau_j$ are Pauli matrices.}
 $$ H_{ST}^{\dagger}\tau_3=\tau_3 H_{ST}. $$
As the Hamiltonian (\ref{ST}) is Hermitian with respect to an inner product $(\Psi_1,\Psi_2)_{\tau_3}=\int \Psi_1^{\dagger}\tau_3\Psi_2 d^3x$, we call it  pseudo-Hermitian \cite{case}. Although a reasonable contribution of the multi-component formalism, it did not resolve the puzzle of the consistent probabilistic interpretation for boson particles. Indefiniteness of the inner product implies existence of wave functions with negative quadrat of the norm. Consequently, continuity equation with indefinite density of probability had to be reinterpreted as a current conservation law, see \cite{FV}.

Recently, several people imagined that the use of the bosonic 
Hamiltonians $H_{ST}$ in relativistic quantum mechanics  may be seen as one of the most
characteristic applications of the so called $PT$-symmetric version
of Quantum Mechanics as proposed by Bender and Boettcher
\cite{BB}. Thanks to a concentrated effort and debate (cf. its
small sample in \cite{ostatni} - \cite{last}) it became clear
that even the operators ${\cal O}$ with the ``weakened" Hermiticity
property
 \be
 {\cal O}^\dagger=\Theta\,{\cal O}\,\Theta^{-1} \neq {\cal O},
 \ \ \ \ \ \ \ \Theta = \Theta^\dagger > 0,
 \ee
(to be called quasi-Hermiticity) may represent, in full accord
with the standard postulates and probabilistic interpretation of
Quantum Mechanics, observable quantities \cite{Geyer}.

The constructive approach of the pseudo-Hermitian quantum mechanics was applied to Klein-Gordon equation \cite{Ali,zno} successfully. In the following article, we intend to perform derivation of the positive metric operator $\Theta$ for the free massive particle with spin one.




The description of our spin-one results will be organized as
follows. Firstly we shall summarize the formulation of a system described by Proca equation. We refresh standard solution of the evolution equation.
We then fix an inertial frame in a way
discussed, originally, by Taketani, Sakata and Tamm \cite{ST}.
This will enable us to outline the construction of the massive
vector boson Hamiltonian (i.e., of the generator of the
time evolution) which appears manifestly non-Hermitian. With the
purpose of achieving an appropriate physical interpretation for the system rewritten in the Schr\"{o}dingerian form we then summarize the application of a few key ideas of
the PT-symmetric Quantum Mechanics to our problem.

Finally, we re-emphasize that the correct
probabilistic interpretation of the Proca's equation must be based
on the introduction and specification of a ``correct" metric
operator $\Theta \neq I$ in the ``physical" Hilbert space of
states. We discuss Lorentz invariance of the associated scalar product.

\section{Proca particle\label{dva}}
\subsection{Covariant formulation}
The equation (\ref{1}) can be rewritten in a more transparent manner without lost of its manifest covariance. Its differentiation and consequent summation
reveal that the individual components of the field have to
fulfill Klein-Gordon equation and are binded by an additional constraint
 \be
 \left(m^2-\Box\right){A}_{\mu}=0,\ \ \ \ \ \ \  \
 \partial_{\nu}{{A}}^{\nu}=0,
  \label{klien}.
  \ee
As could be expected in spin-one system, only three components of the field are
linearly independent.

The free Klein-Gordon equation is solved by
the wave functions
 \be
  A_{\pm}^{\mu}(x)=
 N_{\pm}(p)a^{\mu}_{\pm}(p)e^{-i{p}{x}},
 \ \ \ \ \
 \ \ \ p_0=\pm \omega,
 \ \ \ \ \ \omega=|\sqrt{{\bf p}^2+m^2}|
 \label{3}
 \ee
while the additional condition acquires the form 
 \be p\,a_{\pm}(p)=0.\label{kalibrace}\ee
The three linearly independent four-vectors $a_{\pm}(p)$ may be chosen in the form
 \be
 a_{\pm,1}=(0,{\bf a}_{\pm,1})^T,\ \ \ a_{\pm,2}=(0,
 {\bf a}_{\pm,2})^T
 \ \ \
 a_{\pm,3}=\left(\pm\frac{|{\bf p}|}{\omega},\frac{{\bf p}}{|{\bf p}|}\right)^T
 \label{5}
 \ee
where the space-like components (i.e., polarization vectors) in
the first two items are perpendicular to ${\bf p}$ while the third
one is parallel to ${\bf p}$. For the sake of definiteness we may fix, say,
 \be
 {\bf a}_{\pm,1}({\bf p})=\frac{(0,-p_3,p_2)^T}{\sqrt{p_3^2+p_2^2}},\ 
 \ {\bf a}_{\pm,2}({\bf p})
 =\frac{(p_2^2+p_3^2,-p_1p_2,-p_1p_3)^T}{\sqrt{(p_3^2+p_2^2)^2
 +p_1^2p_2^2+p_1^2p_3^2}}
 \label{7}
 \ee

It is convenient to distinguish the three independent solutions of (\ref{kalibrace}) by their spin properties.  In terms of the three-dimensional representation of the rotation group
 \bea
 S_1=\left(\begin{array}{ccc}0&0&0\\0&0&-i\\0&i&0\end{array}\right),\ \
 S_2=\left(\begin{array}{ccc}0&0&i\\0&0&0\\-i&0&0\end{array}\right),\ \
 S_3=\left(\begin{array}{ccc}0&-i&0\\i&0&0\\0&0&0\end{array}\right)
  \label{reprrot}\eea 
we may define the operator of helicity which in momentum representation reads
 \be
 \hat{h}({\bf p})=\frac{{\bf p}{\bf S}}{|{\bf p}|}={\bf n}{\bf S}=
 i\left(\begin{array}{ccc}0&-n_3&n_2\\n_3&0&-n_1\\-n_2&n_1&0\end{array}\right).
 \label{8}
 \ee
We compose another set of solutions of (\ref{kalibrace}) by a linear combination of the vectors ${ a}_j$
 $$
 {u}_{\pm,1}({\bf p})=\frac{{a}_{\pm,1}({\bf p})+i{a}_{\pm,2}({\bf p})}{\sqrt{2}},\ \
 {u}_{\pm,-1}({\bf p})=\frac{{a}_{\pm,1}({\bf p})-i{a}_{\pm,2}({\bf p})}{\sqrt{2}}
 $$
 \be {u}_{\pm,0}({\bf p})={a}_{\pm,3}({\bf p}).
 \label{helvec}\ee
It can be verified that the spatial components of the four-vectors $u_{\pm,h}$ are eigenstates of the helicity operator (\ref{8})
 \be
 \hat{h}{\bf u}_{\pm,h}({\bf p})=h{\bf u}_{\pm,h}({\bf p}),\ \ h=1,-1,0.
 \label{10}
 \ee
Finally, combining (\ref{3}), (\ref{5}) and (\ref{helvec})  we get
 \be
 A^{\mu}_{\pm,h}({\bf x},t)=N_{\pm,h}u_{\pm,h}^{\mu}({\bf p})
 e^{i{\bf p}{\bf x}\mp i\omega t}
 \label{11},
 \ee
i.e. an explicit form of wave functions with sharp value of energy and helicity in x-representation.

\subsection{Hamiltonian and its wave functions}
As was mentioned in introduction, Feschbach-Villars formulation of integer-spin relativistic equations is closely related to the conception of pseudo-Hermitian operators.
In this section, we refresh Schr\"odinger-like reformulation of the equations (\ref{1}). 
Let us introduce a special denotation of components of antisymmetric tensor $F^{\mu\nu}$ that is in a formal accordance with the one used in electromagnetism
 \bea
 F^{ij}&=&-\varepsilon_{ijm}B_m,\nonumber\\
      F^{0j}&=&-E_j,\ \ i,j=1,2,3.
      \nonumber
      \eea
In a fixed inertial frame, the equations (\ref{1}) then acquire the
form
 \be
 \begin{array}{cclccl}{\bf {B}}&
 =&\rot {\bf {A}},&\f{\partial{\bf A}}{\partial t}&
 =&-{\bf {E}}-\grad A_{0}\\
 A_0&=&-m^{-2}\div{\bf {E}},&\f{\partial{\bf {E}}}
 {\partial t}&=&m^2{\bf {A}}+\rot {\bf {B}}\,.\end{array}
 \label{12}
 \ee
An elimination of ${\bf B}$ and $A_0$ gives
 \bea
 \f{\partial{\bf {A}}}{\partial t}=-{\bf {E}}+\f{\nabla^2{\bf {E}}}{2m^2}
 +\f{\hat{q}}{2m^2}{\bf {E}},\ \
 \f{\partial{\bf {E}}}{\partial t}=m^2{\bf {A}}-\f{\nabla^2{\bf {A}}}{2}
 +\f{\hat{q}}{2}{\bf {A}},
 \label{13}
 \eea
where we abbreviated
 \ben
 \hat{q}=2\grad\div-\nabla^2\,.
 \een
Now, it is easy to arrive at the Schr\"{o}dinger-like matrix equation
 \be
 i\frac{\p}{\p t}\,\Psi =H\Psi,
 \label{ham}
 \ee
where we defined
\be\Psi^T=(m{\bf A},i{\bf E})^T\label{vlnfce}.\ee

The equation (\ref{ham}) prescribes the time-evolution of our system in its
first-quantized form. In the absence of any external forces the
quantum Hamiltonian $H$ of the vector boson is a matrix
containing differential operators as its elements
\footnote{
The Hamiltonian (\ref{brouk}) can be transformed into a rather standard expression for vector boson Hamiltonian (\ref{ST}) wide-spread in the literature (see cf. \cite{ST}, \cite{case}), 
The relation is
 $$ H_{ST}={\cal A}H{\cal A}^{-1},\ \ {\cal A}=i\left(\begin{array}{cc}1&-1\\-1&-1 
 ×
 \end{array}\right)$$
}

 \be
 H=\left(\begin{array}{cc}0&-m+\frac{\grad\div}{m}\\-m
 +\frac{\nabla^2}{m}-
 \frac{\grad\div}{m}&0\end{array}\right).
 \label{brouk}
 \ee
In coordinate representation this manifestly non-Hermitian
$6\times 6-$dimensional matrix has the transparent operator
structure,
 \be
 H=\frac{1}{m}\left(\begin{array}{cccccc}0&0&
 0&\partial_1^2-m^2&\partial_1\partial_2&\partial_1\partial_3\\
 0
 &0&0&\partial_2\partial_1&\partial_2^2-m^2&\partial_2\partial_3\\
 0&0&0&\partial_3\partial_1&\partial_3\partial_2 &\partial_3^2-m^2\\
 -\partial_1^2-\omega^2&-\partial_1\partial_2&-\partial_1\partial_3&0&0&0\\
 -\partial_2\partial_1&-\partial_2^2-\omega^2&-\partial_2\partial_3&0&0&0\\
 -\partial_3\partial_1&-\partial_3\partial_2
 &-\partial_3^2-\omega^2&0&0&0\end{array}\right).
 \label{21,5}
 \ee
This is a non-Hermitian, ${\cal P}-$pseudo-Hermitian operator
generating the free motion,
 \be
 H^{\dagger}={\cal P} H{\cal P}^{-1} \neq H,\ \ \ \ \ \ \ \ \ \ \
  {\cal P}=\left(\begin{array}{cc}0&I\\I&0
  \end{array}\right),
  \label{parita}\ee
where $I$ is $3\times 3$ unit matrix. 

Quite often, the system is considered confined to a finite box of
side length $L$. For the particular periodic boundary conditions
imposed at its walls one then arrives at the ``quantized" energies
on the mass shell,
 \ben
 p_{0,{\bf n}}=\pm \omega_{\bf n}=\pm\sqrt{{\bf p}_{\bf n}^2+m^2},\ \ \ \ \ \
 \ {\bf p}_{{\bf n}}=\frac{2\pi}{L}{\bf n}
 \een
with ${\bf n}$ being a vector of integer components. Vice versa,
the continuity of the energies can be restored in the $L \to
\infty $ limit whenever necessary.
Considering finite $L$ until section \ref{LI}, let us omit the indexes ${\bf n}$. 

To distinguish spin characteristics of wave functions, we introduce another observable which commutes with $H$ in momentum representation 
 \ben
 {\Lambda}_{i,j}({\bf p})={\Lambda}_{i+3,j
 +3}({\bf p})=
 -\frac{i|{\bf p}|
 \varepsilon_{ijk}{p}_k}{{\bf p}^2}=\frac{\bf Sp}{|\bf p|},
 \een
 \be
  {\Lambda}_{i,j+3}({\bf p})={\Lambda}_{i+3,j}
 ({\bf p})=0,\ \ i,j=1,2,3.
 \label{27}
 \ee
This is a block diagonal operator which examines the spin
projection into momentum direction, i.e. it measures the
helicity of our vector boson. It is easy to verify that the latter
operator is also pseudo-Hermitian with respect to ${\cal P}$,
 \ben
 \Lambda^{\dagger}={\cal P}\Lambda{\cal P}^{-1} .
 \een

The six independent eigenvectors of both (\ref{brouk}) and (\ref{27}) may be constructed most easily in terms of the
above-introduced functions (\ref{helvec}) in momentum representation,
 \be
 \Psi_{\pm,h}({\bf p})=\left(\begin{array}{c}
 m{\bf u}_{\pm,h}({\bf p})\\
   \mp \omega{\bf u}_{\pm,h}({\bf p})+{\bf p}u^0_{\pm,h}({\bf p})
             \end{array}\right)
             , \ \ \ \ \ \ \ h=1,-1,0.
             \label{had}\ee
\noindent
We arrive at the complete right-action Schr\"{o}dinger equation
for the fixed-helicity states,
 \ben
  H({\bf p})\Psi_{\pm,h}({\bf p})=\pm\omega\Psi_{\pm,h}({\bf p}),\ \ {\Lambda}({\bf p})\Psi_{\pm,h}({\bf p})=h\Psi_{\pm,h}({\bf p}),\
 .\label{22bia}
 \een
Employing pseudo-Hermiticity of the operators, we can write the relations for their adjoint ones immediately,
 \be
  H^{\dagger}({\bf p}){\cal P}\Psi_{\pm,h}({\bf p})=\pm\omega{\cal P}\Psi_{\pm,h}({\bf p}),\ \ {\Lambda^{\dagger}}({\bf p}){\cal P}\Psi_{\pm,h}({\bf p})=h{\cal P}\Psi_{\pm,h}({\bf p}),\
 \ h=\pm1,\ 0.\label{22bib}
 \ee
It is not difficult to show that the wave functions $\Psi_{\pm,h}({\bf p})$ fulfill the relation
 
\ben
  \langle\Psi_{\epsilon,h}({\bf p}),{\cal P}\Psi_{\epsilon',h'}({\bf p'})\rangle=\delta_{\omega\omega'}\delta_{\epsilon\epsilon'}\delta_{hh'}
  \label{24kj}
 \een
In a way described thoroughly in ref. \cite{Ali}, the set of vectors $\left\{\Psi_{\pm,h}({\bf p}),\ {\cal P}\Psi_{\pm,h}({\bf p})\right\}$ forms a biorthonormal basis of our system in the momentum representation.

\section{Construction of physical metric operator \label{ctyri} }

Let us, for a while, neglect the particular structure of the operators $H$ and $\Lambda$  and consider them just as non-Hermitian operators admitting spectral decomposition in terms of biorthonormal basis \cite{coupledchannels} $\left\{|n,h\rangle,\ |n,h\rangle\rangle|\langle\langle n,h|m,j\rangle=\delta_{mn}\delta_{hj}\right\}$
 \bea H=\sum_n E_n|n,h\rangle\langle\langle n,h|, \ \ 
 \Lambda=\sum_n h|n,h\rangle\langle\langle n,h|\eea
The most general operator ${\cal{P}}$ with respect to which $H$ and  $\Lambda$ are pseudo-hermitian \cite{AM} ($H^{\dagger}={\cal{P}}^{-1}H{\cal{P}}$, $\Lambda^{\dagger}={\cal{P}}^{-1}\Lambda{\cal{P}}$) is
 
\be{\cal P}=\sum_{n,h}|n,h\rangle\rangle \pi_{n,h}\langle\langle n,h|,\ \ \pi_{n,h}\in (-\infty,\infty).\label{genP}\ee

\noindent
Clearly, family of the operators (\ref{genP}) contains positive-definite members. These are just the wanted candidates for a physical metric operator. They are of the form

 \be
 \Theta=\sum_{n,{h}} |n,{h}\rr {\theta}(n,{h}) \ll
 n,{h}|,\ \ {\theta}(n,{h})>0.\label{mys}
 \ee
In the language of ref. \cite{Geyer} the (quasi-)Hermiticity of
$H$ and $\Lambda$ can be recovered by the following redefinition
of the scalar product,
 \ben
 \langle \psi|\phi\rangle_{phys}\equiv\langle \psi|\Theta|\phi\rangle.
 \een
It holds $\langle \psi|H\phi\rangle_{phys}=\langle
H\psi|\phi\rangle_{phys}$, $\langle
\psi|\Lambda\phi\rangle_{phys}=\langle
\Lambda\psi|\phi\rangle_{phys}$. Further examination may proceed
in framework of theory of Hermitian operators where Hilbert space
is generated by rather non-trivial metric $\Theta$ and is spanned
by $\{|n,{h}\rangle\}$.

Refreshing the particular properties of $H $, $\Lambda $ and ${\cal P} $ again, we can rewrite (\ref{mys}) for our system immediately. Indeed, picking up (\ref{parita}), (\ref{had}) and the vectors in (\ref{22bib}) we get
 
\be \Theta_L({\bf p})=\sum_{h}\left[{\theta}(\omega,{h}){\cal
 P}\Psi_{+,h}({\bf p}) \left({\cal P}\Psi_{+,h}({\bf
 p})\right)^{\dagger} +{\theta}(-\omega,{h}){\cal P}\Psi_{-,h}({\bf
 p}) \left({\cal P}\Psi_{-,h}({\bf p})\right)^{\dagger}\right],\nonumber\ee 
\be \theta(\pm,h)>0,\ h=\pm 1,0.  \label{25}\ee

\noindent
Performing quite straightforward calculations, we arrive to the
following expression
  \bea
  \Theta_L({\bf p})&=&\left(\begin{array}{cc}
  \Theta({\bf p})_{11}&\Theta({\bf p})_{12}\\
  \Theta({\bf p})_{12}&\Theta({\bf p})_{22}
  \end{array}\right).\label{29}
  \eea
\noindent
Considering the explicit form of the involved quantities (\ref{5}), (\ref{7}), (\ref{helvec}) and (\ref{had}), it could be expected that the resulting matrix-operators $\Theta_{ij}$ will be rather complicated. However, these revealed to be of an unexpectedly simple structure

\bea
 \Theta({\bf p})_{11}&=&s_{+,1}J^T({\bf p})+s_{+,-1}J({\bf p})+
 s_{+,0}K({\bf p})\nonumber \\
 \Theta({\bf p})_{12}&=&\frac{m}{\omega}\left[s_{-,1}J^T({\bf p})+
 s_{-,-1}J({\bf p})
  +\frac{\omega^2}{m^2}s_{-,0}K({\bf p})\right]\nonumber\\
 \Theta({\bf p})_{22}&=&\frac{m^2}{\omega^2}\left[s_{+,1}J^T({\bf p})+
 s_{+,-1}J({\bf p})+\frac{\omega^4}{m^4}s_{+,0}K({\bf p}) \right], \nonumber\\
s_{\pm,h}&=&\pm{\theta}(\omega,h)+{\theta}(-\omega,h)
 , \label{29,5}
 \eea

\noindent
in terms of $3\times 3$ Hermitian matrices $J$ and $K$ with
elements
 \be
 J({\bf p})_{i,j}=\frac{{\bf p}^2\delta_{ij}-p_ip_j
 +i\varepsilon_{ijk}p_k|{\bf p}|}{8m^2{\bf p}^2},\ \ \
   K({\bf p})_{i,j}=\frac{p_ip_j}{4m^2{\bf p}^2},
    \ \ i<j,\ j=1,2,3.
    \label{29,7}
    \ee
Moreover, the matrices $J$ and $K$ can be expressed as polynimials of the
helicity operator (\ref{8})
 \be
 J=\frac{\hat{h}^2-\hat{h}}{8m^2},\
  \ K=\frac{I-\hat{h}^2}{4m^2}.
  \ee
Then, the metric components (\ref{29,5}) reveal to be functions of
energy and helicity operators. The expression (\ref{29,5}) represents the candidate for proper metric operator.

However, it suffers from a strong ambiguity represented by the six unknown functions $\theta(\pm\omega,h)$. In the following section, we will impose an additional physically motivated requirement which will reduce the unwanted degrees of freedom. 

%
%
%


\section{Lorentz invariance\label{LI}}
Let us send the length $L$, determining the volume in which the particle is confined, in infinity. 
Then the problem appears how to describe the system from various inertial frames of reference. To resolve the puzzle, it is crucial to find an appropriate representation of Poincare group on a space spanned by wavefunctions. 

It is an immediate physical requirement that the scalar product should be independent on the reference frame. Writing this condition explicitly, we have
  \be (\Psi'_1,\Theta\Psi_2')=(\Psi_1,\Theta\Psi_2),\label{invofscalprod}\ee
where $\Psi'_j$ are transformed wavefunctions. In standard case where $\Theta=1$, this condition is ensured by unitary representation of Poincare group on the Hilbert space of states. 

Let us consider infinitesimal transformations of the wavefunctions
 \be\Psi'({\bf x},t)=(1+i\epsilon M)\Psi({\bf x},t),\ \ \epsilon\sim 0\label{infinitrans}\ee
where $M$ is a generator of Poincare algebra in an appropriate representation. 
Then the invariance of scalar product leads to the operator equation
 \be M^{\dagger}\Theta=\Theta M.\label{lepoun}\ee
It can be expected that this will reduce the free parameters in the metric operator additionally.

To handle with the relation (\ref{lepoun}), we need to find an explicit form of generators of Poincare group. 
As the transformation properties of wavefunctions (\ref{11}) are known since Wigner \cite{wigner}, it is possible to derive the transformation law for (\ref{vlnfce}) directly. Instead of performing this straightforward but rather lengthy computation, it is sufficient to refer the literature.
In 1958, Foldy encountered the problem for relativistic systems with various spins \cite{foldy} (see also \cite{Nikitin}). Let us follow the existing procedure by introducing a transformation\footnote{It has been found by Case \cite{case} as a generalization of Foldy-Wouthuysen transformation to spin one.} 
\be {\cal B}={(8\omega)}^{-1/2} \left( \begin{array}{cc} 
 \omega +m+\frac{q}{\omega +m}&-\omega -m+\frac{q}{\omega +m}\\ -\omega -m-\frac{q}{\omega +m}&-\omega -m+\frac{q}{\omega +m}\end{array}
\right), \chi = {\cal B} \Psi,\label{foldytr2}\ee
where $\Psi$ is six component wave function (\ref{vlnfce}). The transformation brings the Hamiltonian $H$ in (\ref{brouk}) into the diagonal form
 
 \be H_F={\cal B}H{\cal B}^{-1}=\omega\beta,\ \ \
	\ee
while the operator $\Lambda$ remains unchanged under this transformation.
Foldy found that the generators of infinitesimal transformations of Poincare group are
 \bea 	H_F&=&\beta\omega,\nonumber\\
	{\bf P}&=&{\cal I}{\bf p},\nonumber\\
	{\bf K}&=&\frac{\beta}{2}({\bf x\omega+\omega {\bf x}})-\beta\frac{{\bf S}\times{\bf p}}{m+\omega}-{\cal I}t{\bf p}, 
	\nonumber\\
	{\bf J}&=&{\cal I}\left({\bf x}\times {\bf p}+{\bf S}\right),\ \ \beta=\left(\begin{array}{cc}I&0\\0&-I
	\end{array}\right) ,\ \ {\cal I}=\left(\begin{array}{cc}I&0\\0&I
	\end{array}\right).\ \ \label{foldytr1}
\label{generatory}
      \eea 
\noindent
These constitute a direct sum of two irreducible representations of Poincare algebra\footnote{For explicit form of discrete symmetries see \cite{foldy}}. They are hermitian with respect to the scalar product
 \be (\chi_1,\chi_2)=\int \chi_1^{\dagger}\chi_2d^3x.\label{beznysoucin}\ee
The representation (\ref{generatory}) is called Foldy`s canonical representation and appears in many works considering relativistic equations of free particles \cite{fonda}, \cite{mathews}.

As we require ${\cal B}$ to be a unitary mapping the new metric operator reads $\tilde{\Theta}=({\cal B}^{-1})^{\dagger}\Theta {\cal B}^{-1}$
and is block diagonal\footnote{The initial indefinite metric $ {\cal P}$ appearing in (\ref{parita}) revealed to be proportional to a proper space inversion 
 $$({\cal B}^{-1})^{\dagger}{\cal P B}^{-1}=\frac{1}{m}
	\left(\begin{array}{cc}-I&0\\0&I\end{array}\right),
 $$

\noindent 
discrete transformation within the Poincare group (see \cite{foldy}).}
 \be\tilde{\Theta}=\left(\begin{array}{ll}\tilde{\Theta}_{+}&0\\0&\tilde{\Theta}_-\end{array}
\right).\nonumber\ee
The diagonal sub-matrices-operators can be written in a compact form
 $$\tilde{\Theta}_{\pm} = \frac{1}{4m^2\omega}\left[\left(\theta(\pm\omega,1)+\theta({\pm\omega,-1})-
\frac{2\omega^2}{m^2}\theta(\pm\omega,0)\right)\hat{h}^2\right.$$
\be\left.+\left(\theta(\pm\omega,1)-\theta(\pm\omega,-1)\right)\hat{h}+\frac{2\omega^2}{m^2}\theta(\pm\omega,0)I\right].\label{tucnak}\ee
Now, let us return to the Lorentz invariance condition (\ref{lepoun}). The generators (\ref{generatory}) are hermitian with respect to (\ref{beznysoucin}). Thus, we can rewrite the condition as

 $$M\tilde{\Theta}=\tilde{\Theta}M\Longrightarrow [M,\tilde{\Theta}]=0. $$
\noindent
It is a well known result of representation theory that the operator commuting with all generators of an irreducible representation is proportional to identity on the representation space. In our case, we deal with reducible representation of the Poincare algebra.
Consequently, not the metric operator $\tilde{\Theta}$ but its particular components $\tilde{\Theta}_{\pm}$ have to be multiples of the identity operator. Considering their explicit forms (\ref{tucnak}), we arrive to the set of relations for unknown functions $\theta(\pm\omega,j)$
 \bea	\theta(\pm\omega,1)+\theta(\pm\omega,-1)&=&\frac{2\omega^2}{m^2}\theta(\pm\omega,0)\nonumber\\
	\theta(\pm\omega,1)&=&\theta(\pm\omega,-1)\nonumber\\
	\omega\theta(\pm\omega,0)&\sim&1\label{rce}.\eea
Their solution restrict ambiguity of the metric operator drastically since we obtain

\be \theta(\pm\omega,0)=\frac{\alpha_{\pm}m}{\omega},\ \ \theta(\pm\omega,j)=\frac{\alpha_{\pm}\omega}{m},\ \ j=\pm 1\label{parametry},\ee

\noindent
where $\alpha_{\pm}$ are dimensionless positive real numbers. 

The evolution equation in Foldy`s canonical form is $i\partial_t \chi=H_f\chi$. It is homogenous, i.e. its two solutions which differ by multiplication constant are physically equivalent. Thus, one of the parameters $\alpha_{\pm}$ can be immersed into a wave function.  Concluding, the most general form of the positive definite, Lorentz invariant metric operator for free spin-one particle in Foldy`s canonical representation is  
\footnote{It can be transformed, up to a multiplicative constant, into a diagonal matrix. 
  \be ({\cal D}^{-1})^{\dagger}\tilde{\Theta}{\cal D}^{-1}=\frac{1}{m^3}\left(\begin{array}{cc}1&0\\ 0& 1 \end{array}\right),\ \
	{\cal D}=\left(\begin{array}{cc}1&0\\ 0&\sqrt{\gamma} \end{array}\right).\nonumber \ee 
\noindent
The remaining free parameter is eliminated from the metric. However, probabilistic interpretation remains non-unique as the ambiguity moves to the definition of wavefunction.}

 \be\tilde{\Theta}=\frac{1}{m^3}\left(\begin{array}{cc}I&0\\0&\gamma I
  \end{array}
 \right), \ \ \gamma=\frac{\alpha_-}{\alpha_+}.\label{SSFoldy}\ee

Finally, let us leave the canonical formalism and return to the space of solutions (\ref{11}) of Proca equation (\ref{1}). We can state the following\\
\textit{Theorem: Positive-definite and Lorentz invariant scalar products on the space of solutions of Proca equation (\ref{1}) form a one-parameter set. Explicitly, the scalar product of two solutions of Proca equation $A_j=(A_j^0,{\bf A}_j)^T,\ j=1,2$  is}

 \be((A_1,A_2))\equiv-\frac{i\alpha}{2m}
	\left[
		\left\langle{\bf A}_1,\dot{{\bf A}}_2+\grad A_2^0\right\rangle
	       -\left\langle\dot{{\bf A}}_1+\grad A_1^0,{\bf A}_2\right\rangle
	\right]
 \nonumber\ee
 \be+
	\left\langle
		{\bf A}_1,\left[\frac{m}{\hat{\omega}}I+\frac{\left({\bf pS}\right)^2}{2m\hat{\omega}}\right]{\bf A}_2\right\rangle
 +	\left\langle
		\dot{{\bf A}}_1+\grad A^0_1,\left[\frac{\hat{\omega}}{2m^3}I-\frac{\left({\bf pS}\right)^2}{2m^3\hat{\omega}}\right]\left(\dot{{\bf A}}_2+\grad A_2^0\right)\right\rangle\label{theorem}\ee
\textit{where $\alpha$ is in $(-1,1)$, $\langle.,.\rangle$ denotes standard scalar product and $\hat{\omega}=\int d{\bf  p}\omega|{\bf p}\rangle\langle {\bf p}|$.}\\
\noindent
\textit{Proof:}

Let us recall the definition of $\Psi=(m{\bf A},-i\partial_t {\bf A}-i\grad{A^0})^T$ and get  the reverse transformation of $\tilde{\Theta}$
 $$\Theta={\cal B}^{\dagger}\tilde{\Theta}{\cal B}=\frac{\alpha_--\alpha_+}{2m^2}
\left(\begin{array}{cc}0&I\\I&0\end{array}\right)+
\frac{\alpha_-+\alpha_+}{2m^3}\left(\begin{array}{cc}\frac{m^2}{\hat{\omega}}I+\frac{\left({\bf pS}\right)^2}{\hat{\omega}}&0\\
	0&\hat{\omega} I-\frac{\left({\bf pS}\right)^2}{\hat{\omega}}\end{array}\right).
 $$
As Proca equation (\ref{1}) is homogenous, we can  absorb one of the coefficients, say $\alpha_++\alpha_-$, to the wave function. Inserting these quantities into 
 $$((A_1,A_2))\equiv (\Psi_1,\Theta\Psi_2).$$
and denoting $\alpha\equiv \alpha_--\alpha_+$, we arrive to the expression (\ref{theorem}).


\section{Discussion}\label{sest}

In the present article, we considered a free massive vector boson.
After a short introduction into the its standard quantum description, we re-derived its description within multicomponent Hamiltonian formalism. Consequently, we
summarized main concepts and results concerning pseudo-Hermitian
diagonalizable operators with discrete spectra and applied them to
the quantum system. As the dynamics of free massive spin-one particle can be rewritten via apparently
non-Hermitian operators~(\ref{brouk}),  we aimed to apply
framework of pseudo-Hermitian quantum mechanics to this quantum
relativistic system. Particularly, we constructed a positive
definite operator that plays the role of the positive definite
metric on the space of solutions of evolution equation
(\ref{ham}). 

Its construction proceeded in two steps. First, we considered the particle to be confined in a box with finite length $L$ of side. Then, the resulting positive-definite metric operator acquired the form (\ref{29,5}). It revealed to be a function of energy and helicity operators. It had six degrees of freedom represented by unknown functions $\theta(\pm,j),\ j=\pm 1,0$. 
 
Subsequently, we made a transition to the open space by sending $L$ to infinity. We discussed Lorentz invariance of the metric operator (\ref{29,5}). This physically motivated requirement reduced the ambiguity of the metric. We found, that the Lorentz invariant positive-definite scalar products (\ref{SSFoldy})  create a one parameter family. We found an explicit form of scalar product for solutions of Proca equation (\ref{theorem}) which is in a close analogy to its spin-zero counterpart derived by Mostafazadeh \cite{Ali}. In this way, we contributed to the probabilistic interpretation of the Proca wave functions.

At the very end, let us still make a few remarks. Attempts to construct a proper probabilistic interpretation within the relativistic quantum mechanics are not new. To our best knowledge, the most explicit and the oldest work in this direction was done by Mathews \cite{mathews}. In comparison with the approach of the present article and \cite{Ali}, he enquired the problem form the very opposite side. He started the construction with a known representation of Poincare algebra for free particles. He seeked an operator which would generate Lorentz invariant scalar product. Although not required from the beginning, the resulting operator was positive-definite. Pseudo-Hermitian quantum mechanics enables construction of the metric operator in relativistic quantum systems where the generators of Poincare group are not given explicitly. In this sense, our approach can be seen as a more general.  

\subsection*{Acknowledgement}
Participation of VJ  was partially supported by the project LC06002.

\end{document}